\documentstyle[aps,amssymb,prl,preprint]{revtex}
\begin{document}
\draft         
\author{I.V. Barashenkov\footnote{Supported by a visiting fellowship
from ICTP, Trieste}\footnote{Email: igor@uctvms.uct.ac.za}
 and
D.E. Pelinovsky\footnote{
Now at Department of Mathematics, University of Toronto,
100 St.George street, Toronto, Ontario, Canada M5S 3G3.}\footnote{
Email: dmpeli@math.toronto.edu}
}

\address{Department of Mathematics, University of Cape Town,
 Rondebosch 7701, South Africa}

\title{Exact vortex solutions of the complex sine-Gordon theory
on the plane}

\date{\today}
\maketitle

\begin{abstract}
We construct explicit multivortex solutions for the first and
second complex sine-Gordon equations. The constructed solutions
are expressible in terms of the modified Bessel and rational
functions, respectively. The vorticity-raising and lowering
B\"acklund transformations  are
interpreted as the Schlesinger transformations of the fifth
Painlev\'e equation.
\end{abstract}

\vspace{10mm}


\vspace{1cm}

\newpage

\noindent
{\it Motivation.} \, Recently there has been an
upsurge of interest in the
 complex sine-Gordon equation.
Originally derived in the
reduction of the $O(4)$ nonlinear $\sigma$-model \cite{Pohlmeyer} and
a theory of dual strings interacting through a scalar field
\cite{Lund_Regge},  this equation  reappeared in a number of
field-theoretic
 \cite{Neveu_Papanicolaou}
and   fluid dynamical \cite{Fukumoto} contexts.
The
equation was shown to be completely integrable
\cite{Pohlmeyer,Lund,G1},
and the multisoliton solutions were constructed in a variety of
forms, both over vanishing \cite{dVM2,BG1} and nonvanishing
backgrounds \cite{G2,BG2}. The study of its quantized version
started in \cite{dVM2,dVM1,BD} and received a new impetus
recently \cite{recent}  when it was realized that the
complex sine-Gordon theory may be reformulated in terms
of the gauged Wess-Zumino-Witten action and interpreted as an 
integrably
deformed SU(2)/U(1)-coset model \cite{Bakas_Park}.

The complex sine-Gordon theory can be conveniently defined by its
action functional,
\begin{equation}
\label{E_cSG1}
E_{SG-1} = \int \left[ |\nabla \psi|^2 +
(1 - |\psi|^2)^2  \right] \frac{d^2x}{1-|\psi|^2}.
\end{equation}
The  subscript 1 serves to distinguish this model from
another  integrable complexification of
the sine-Gordon theory, the so-called complex sine-Gordon-2:
\begin{equation}
\label{E_cSG2}
E_{SG-2} = \int \left[ \frac{|\nabla \psi|^2}{1-\frac{1}{2} |\psi|^2}
+ \frac{1}{2} (1 - |\psi|^2)^2  \right] {d^2x}.
\end{equation}
The latter system was derived in ref.\cite{Sciuto} as the bosonic 
limit of a
generalized supersymmetric sine-Gordon equation and, independently,
in ref.\cite{Get_csG-II}. Quantum mechanically, the above two complex
sine-Gordon models were shown to be the only O(2)-symmetric theories
whose $S$-matrix is factorizable at the tree level\cite{BD}.

In all previous analyses the complex sine-Gordon equations were 
considered
in the
(1+1)-dimensional Minkowski space-time. In the present Letter we
study these two models  in the
2-dimensional Euclidean space.
One reason for this is that
they define integrable
perturbations of  Euclidean conformal field
theories; more precisely,
eqs.(\ref{E_cSG1})-(\ref{E_cSG2})
arise as reductions of the  $SU(2)_N$ gauged Wess-Zumino-Witten model
perturbed by a multiplet of primary fields
(by $\Phi^{(1)}$ and $\Phi^{(2)}$, respectively) 
\cite{Bakas_Park,Brazhnikov}.
They  are closely related to  important two-dimensional
lattice  systems, viz.
 $Z_N$ parafermion models perturbed
by the first and second thermal operators, respectively \cite{Fateev}.

Another motivation for studying solutions of the
Euclidean complex sine-Gordon equations comes from
 a remarkable similarity between eqs.(\ref{E_cSG1})-(\ref{E_cSG2})
and several phenomenological lagrangians of condensed matter physics,
in particular the
Ginsburg-Landau expansion  of the free energy in the theory of  phase 
transitions,
\begin{equation}
\label{GL}
E_{GL} = \int \left[ |\nabla \psi|^2 + \frac{1}{2} (1 - |\psi|^2)^2
\right] d^2x,
\end{equation}
and the energy of the Heisenberg ferromagnet with easy-plane 
anisotropy \cite{Magnetism}:
\begin{equation}
E_{FM}=
\int \left[ (\nabla \alpha)^2 + \sin^2 \alpha (\nabla \beta)^2
+ \cos^2 \alpha \right] d^2x.
\label{FM}
\end{equation}
(Hence we will be using the words ``action" and ``energy"
interchangeably in what follows.)
To see that
(\ref{E_cSG1})-(\ref{E_cSG2}) are relatives of (\ref{FM}), one
writes $\psi= \sin \alpha e^{i \beta}$ and $\psi=
\sqrt{2} \sin (\alpha/2) e^{i \beta}$, transforming
eqs.(\ref{E_cSG1}) and
(\ref{E_cSG2})  into
\begin{equation}
E_{SG-1}=
\int \left[ (\nabla \alpha)^2 +  \tan^2 \alpha (\nabla \beta)^2
+ \cos^2 \alpha \right] d^2x
\label{a2}
\end{equation}
and
\begin{equation}
E_{SG-2}= \frac 12
\int \left[ (\nabla \alpha)^2 + 4 \tan^2 \frac{ \alpha }{2} (\nabla 
\beta)^2
+ \cos^2\alpha \right] d^2x,
\label{a3}
\end{equation}
respectively.

The Ginsburg-Landau free energy (\ref{GL}) is minimized by
the  Gross-Pitaevski  vortices
originally discovered in the context of
superfluidity \cite{superfluidity}. These are
topological solitons of the form $\psi(x,y)=\Phi_n(r)e^{in \theta}$,
where 
$\Phi_n \to 1$ as $r \to \infty$. Although these
important solutions were obtained numerically  and in
various asymptotic regimes, no analytic expressions
for the Gross-Pitaevski vortices are available.
Similarly, eq.(\ref{FM}) is minimized by magnetic vortices
\cite{Magnetism}, and again, these are available only numerically.
The aim of this note is  to demonstrate that
 the Euclidean complex sine-Gordon
 equations also exhibit
topological soliton solutions. Unlike the Gross-Pitaevski vortices
and unlike their magnetic counterparts,
the vortices of eqs.(\ref{E_cSG1}) and (\ref{E_cSG2}) can be found
exactly, and in a closed analytic form.
Consequently, the significance of the complex sine-Gordon equations
on the plane stems from the fact that they provide a laboratory for
studying analytic properties  of vortices
and their phenomenology in
a wide class of condensed matter models.

 We construct these solutions
in two different ways: (i) by means of an auto-B\"acklund
transformation
resulting from the spinor representation of the complex sine-Gordon
theory, and (ii) via the Schlesinger transformation of the fifth
Painlev\'e equation,
\begin{eqnarray}
 W_{rr} +\frac 1 r W_r - \left( \frac 1{W-1} +
\frac 1 {2W} \right) W_r^2 = \frac{(W-
1)^2}{r^2}
\left( \alpha W + \frac{\beta}{W} \right) + \frac{\gamma}{r} W +
\delta W \frac{W+1}{W-1},
\label{P5}
\end{eqnarray}
which arises in a self-similar reduction of eqs.(\ref{E_cSG1})
and (\ref{E_cSG2}).

{\it Vortices via B\"acklund transformation.} \,
The complex sine-Gordon-1 equation,
\begin{equation}
\nabla^2 \psi + \frac{(\nabla \psi)^2 \, {\overline \psi}}{1- 
|\psi|^2}
+ \psi(1-|\psi|^2)=0,
\label{csG-I}
\end{equation}
\begin{mathletters}
admits an equivalent representation in terms of the Euclidean
spinor field, $\Psi=(u,v)^T$ \cite{BG2}:
\begin{eqnarray}
\label{MTM2}
i \, {\overline \partial} u   + v - |u|^2  v & = & 0,\\
\label{MTM1}
i \, \partial v +  u - |v|^2  u & = & 0.
\end{eqnarray}
\label{MTM}
\end{mathletters}[Here $\partial = \partial/\partial z$,
${\overline \partial} = \partial/ \partial {\overline z}$
and $z=(x+iy)/2$.] This is nothing but the Euclidean version
of the massive Thirring model; the corresponding action
functional has the form
\begin{eqnarray}
E_{Th}= \int \left[ i \Psi^{\dag} \gamma_i \partial_i \Psi
+ \Psi^{\dag} \Psi - \frac14
\left( \Psi^{\dag} \gamma_i \Psi \right)^2 -1 + c.c.\right] d^2x
\nonumber\\
= \int  \left( i {\overline u} \, \partial v + i {\overline v}
\, {\overline \partial} u + |u|^2 + |v|^2 - |uv|^2 -1 + c.c.
\right) d^2x.
\label{Th_action}
\end{eqnarray}
Since as one can easily check
both $u$ and $v$ satisfy eq.(\ref{csG-I}), the
 Thirring model (\ref{MTM}) can be regarded as a
B\"acklund transformation  between two different solutions of
eq.(\ref{csG-I}). Here we confine ourselves to multivortex solutions
of the form $\psi = \Phi_n(r) e^{in \theta}$,
where ($r,\,\theta$) are polar coordinates on the plane and
$\Phi_n(r)$ is a real function satisfying
\begin{equation}
\label{ODE1}
\frac{d^2 \Phi_n}{d r^2} + \frac{1}{r} \frac{d \Phi_n}{d r} +
\frac{\Phi_n}{1 - \Phi_n^2} \left[ \left( \frac{d \Phi_n}{d r}
\right)^2
- \frac{n^2}{r^2} \right] + \Phi_n \left( 1 - \Phi_n^2 \right) = 0.
\end{equation}
Eqs.(\ref{MTM}) with $u$ and $v$ of the form
$u=-i\Phi_{n-1}e^{i(n-1)\theta}$
and $v= \Phi_n
e^{i n \theta}$
 furnish an
equivalent representation for eq.(\ref{ODE1}):
\begin{mathletters}
\label{sys}
\begin{eqnarray}
\label{sys2}
-\frac{d \Phi_{n-1}}{d r} +\frac{n-1}{r} \Phi_{n-1} & = &
 ( 1 - \Phi_{n-1}^2 ) \Phi_n,\\
\label{sys1}
\frac{d \Phi_n}{d r} + \frac{n}{r} \Phi_n & = &
( 1 - \Phi_n^2) \Phi_{n-1},
\end{eqnarray}
\end{mathletters}where
 $\Phi_n$ and $\Phi_{n-1}$ satisfy eq.(\ref{ODE1}) with
$n$ and $n'=n-1$, respectively. When $n=1$, eq.(\ref{sys2}) is solved
by $\Phi_0=1$ and eq.(\ref{sys1}) becomes a Riccati equation:
\begin{equation}
\label{Riccati1}
 \Phi_1' + r^{-1} \Phi_1 = 1 - \Phi_1^2.
\end{equation}
This equation can be linearized by writing $\Phi_1 =  S'/S$,
where $S(r)$ satisfies the modified Bessel's equation of zero order:
$S''+S'/r-S=0$. Selecting $S = I_0(r)$ gives the explicit form of
the $n=1$ vortex solution of the complex sine-Gordon theory:
\begin{equation}
\label{fund}
\Phi_1 = \frac{I_1(r)}{I_0(r)}.
\end{equation}
Here $I_0(r)$ and $I_1 = I_0^{\prime}(r)$ are the modified Bessel
functions of zero and first order, respectively. The vortex is
plotted in Fig.1.

With the solution $\Phi_1$ at hand, eqs.(\ref{sys}) yield a recursion
relation allowing us to construct solutions with vorticity $n>1$ in
a purely algebraic way:
\begin{equation}
\label{recur}
\Phi_{n+1} = \frac{-1}{1 - \Phi_n^2} \left[ \frac{d \Phi_n}{dr} -
\frac{n}{r} \Phi_n \right] = \Phi_{n-1} - \frac{2}{1 - \Phi_n^2}
\frac{d \Phi_n}{dr},\;\;\;\;n \geq 1.
\end{equation}
In particular, the first two higher-order vortices (shown in Fig.1)
are given by
\begin{eqnarray*}
\Phi_2 & = & - \frac{I_0 I_2 - I_1^2}{I_0^2 - I_1^2}, \\
\Phi_3 & = & \frac{(I_3 - I_1)(I_0^2 - I_1^2) + I_1 (I_0 - I_2)^2}{
(I_0 - I_2)(I_0 I_2 - 2 I_1^2 + I_0^2)},
\end{eqnarray*}
where we have eliminated derivatives by means of the well known
relation between the modified Bessel functions of different order:
$I_{n+1} + I_{n-1} = 2 I_n'$.
The
asymptotic behaviour
of the vortex with vorticity $n$ is readily found
from eq.(\ref{sys1}):
\begin{eqnarray}
\label{asympt1}
\Phi_n & \sim & \frac{1}{2^n n!} r^n - \frac{1}{2^{n+2} (n+1)!}
r^{n+2} + {\rm O}(r^{n+4})\;\;\;\;{\rm as}\;\;\;\;r \rightarrow 0, \\
\Phi_n & \sim & 1 - \frac{n}{2r} - \frac{n^2}{8 r^2} +
{\rm O}(r^{-3})\;\;\;\;{\rm as}\;\;\;\;r \rightarrow \infty.
\label{asympt2}
\end{eqnarray}
One consequence of eq.(\ref{asympt2}) is that the energy of the
vortices diverges [cf. eq.(\ref{below}) below], similarly to the 
energy
of
the Gross-Pitaevski and easy-plane ferromagnetic vortices
\cite{superfluidity,Magnetism}. (Physically, this fact simply 
indicates
that there is a cut-off radius in the system, for example
the radius of the cylindrical superfluid container, or the distance
between two adjacent vortex lines.)

{\it Bogomol'nyi bound.} \,
An important question is whether the vortex renders the action
a minimum. Let $n=1$ and rewrite eq.(\ref{E_cSG1}) as
\begin{equation}
\label{Bogomolny}
E_{SG-1} =  \int \left| \partial \psi  +|\psi|^2 -1 \right|^2
\frac{d^2x}{1-|\psi|^2} + \int \nabla \cdot {\bf A} d^2x,
\end{equation}
where ${\bf A}$ is a real vector field with components
\[
A_i =
\ln( 1 - |\psi|^2)  \,
\epsilon_{ij} \partial_j
  {\rm Arg} \, \psi+ 2\psi_i; \quad i=1,2,
\]
and $\psi = \psi_1 + i \psi_2$.
Assume our fields are such that $|\psi|^2 <1$; then the first term
in (\ref{Bogomolny}) attains its minimum at
solutions to the ``Bogomol'nyi equation" $\partial \psi=1-|\psi|^2$.
This is exactly our eq.(\ref{MTM1}) with $v=\psi$ and $u=-i$;
its vortex solution is given by eq.(\ref{fund}).
The
second integral in (\ref{Bogomolny}) represents the divergent part
of the action; it
can be written as a flux through a
circle of the radius $R \to \infty$.
 Perturbing the vortex
$\psi = \Phi_1(r) e^{i \theta}$ by a function $\delta \psi$
decaying faster than $1/r$ at infinity will not affect this part;
the flux is uniquely determined by the vortex asymptotes:
\begin{equation}
\label{below}
\oint_{C_R} {\bf A} \cdot {\bf n} \, dl= 2 \pi (2R- \ln R-1) + {\cal
O}(R^{-
1}).
\end{equation}
Consequently, the $n=1$ vortex saturates the minimum of the action
in the class of functions with $|\psi|^2<1$.

The importance of the last inequality should be specially
emphasized.
 Without the
condition $|\psi|^2<1$
 being imposed, one could construct a perturbation
${\tilde \psi}(x,y)$ of the vortex satisfying $|{\tilde \psi}|=1$,
$\nabla {\tilde \psi}=0$ on some closed curve
on the $(x,y)$-plane which does not
  enclose
the origin. Taking then $|{\tilde \psi}| \gg 1$
in the interior of this contour, the action (\ref{E_cSG1})
could  be made arbitrarily negative.

It is interesting to note that the first-order equations
(\ref{MTM}) with generic $u$ and $v$ can also be interpreted
as the Bogomol'nyi limit for some more general system with
twice as many degrees of freedom. The corresponding action
functional is
\begin{eqnarray}
E[u,v]
= \int \left( \frac{|\nabla u|^2}{1- |u|^2} +
1 - |u|^2  \right) d^2x
+ \int \left( \frac{|\nabla v|^2}{1- |v|^2} +
1 - |v|^2  \right) d^2x + E_{Th},
\label{general}
\end{eqnarray}
where $E_{Th}$ is the Thirring action (\ref{Th_action}). Clearly, any
solution to eq.(\ref{MTM}) is automatically a solution to the
second-order system (\ref{general}). The action (\ref{general})
can be written as
\begin{eqnarray}
E[u,v]
= \int \frac{| i \, {\overline \partial}  u + v(1-|u|^2)|^2}{1- |u|^2}
 d^2x
+ \int \frac{| i \, { \partial}  v + u(1-|v|^2)|^2}{1- |v|^2}d^2x
+ \int \nabla \cdot {\bf A} d^2x,
\label{gene_Bogo}
\end{eqnarray}
where $A_i=\ln (1-|v|^2) \epsilon_{ij} \partial_j {\rm Arg\/}v-
\ln (1-|u|^2) \epsilon_{ij} \partial_j {\rm Arg\/}u$.
Assuming, again, that $|u|^2, |v|^2<1$, the lower bound of
the action (\ref{general}-\ref{gene_Bogo}) is saturated by solutions 
to
eqs.(\ref{MTM}).

Some properties of the complex sine-Gordon vortices receive a
natural interpretation when the equation is reformulated as
a $\sigma$-model on a two-dimensional surface $\Sigma$ embedded
in a three-dimensional space $(n_1, n_2, n_3)$. The metric on $\Sigma$
is $ds^2=d \alpha^2 + {\rm tan}^2{\alpha}\, d \beta^2$ [see 
eq.(\ref{a2})].
In order for $\Sigma$ to be smooth, the space $(n_1,n_2, n_3)$
has to be pseudoeuclidean and the surface noncompact; in fact it
looks like an asymptotically conical infinite bowl:
$$ n_1+ in_2= {\rm tan} \alpha \, e^{i \beta}; $$
$$n_3= \frac1q - {\rm tanh}^{-1} q, \quad
q= \frac{\cos  \alpha}{(1+ \cos^2  \alpha)^{1/2}}. $$
Here $0 \leq \alpha < \pi/2$, $0 \leq \beta < 2 \pi$. In terms of
$n_i$, the lagrangian (\ref{E_cSG1}) reads
$$ E_{SG-1}= \int  \left[
(\nabla n_1)^2 + (\nabla n_2)^2 - (\nabla n_3)^2+
(1+ n_1^2 + n_2^2)^{-1} \right] d^2x. $$
As $r \to  \infty$, all three components of the vortex field,
$n_1$, $n_2$ and $n_3$, tend to infinity.  Consequently, the vortices
map a noncompactified $(x,y)$-plane onto a noncompact surface ---
this accounts for their infinite energy. We also acknowledge the
role of the
condition $|\psi|^2<1$, which characterizes solutions
admitting the $\sigma$-model interpretation.

{\it Reductions to the Painlev\'e-V.} \,
The transformation
$$
\Phi_n = \frac{1 + W}{1 - W}
$$
reduces eq.(\ref{ODE1}) to the
fifth Painlev\'e equation (\ref{P5})
 with coefficients
\begin{equation}
\alpha = {n^2}/{8},\;\;\;\;\beta = - {n^2}/{8},\;\;\;\;
\gamma = 0,\;\;\;\;\delta = -2.
\label{coefficients}
\end{equation}
For $\gamma=c(1-a-b)$, where $a^2=2 \alpha$, $b^2= -2 \beta$,
and $c^2=-2 \delta$, eq.(\ref{P5}) admits a reduction
\cite{Luk} to a Riccati equation
\begin{equation}
\label{bel1}
W_r = r^{-1} (W - 1)(a W + b) + c W.
\end{equation}
The above relation between
the coefficients is in place for $n=1$; in terms of the vortex modulus
 $\Phi_1$, eq.(\ref{bel1}) turns out to be nothing but
our eq.(\ref{Riccati1}).
Next, the Schlesinger transformations of the Painlev\'e-V
to itself \cite{Grom,Fok} have the form
\begin{mathletters}
\label{bel2}
\begin{eqnarray}
W_r & = & r^{-1} (W - 1)(a W + b) + c W \frac{1 +
\hat{W}}{1 - \hat{W}}, \\
-  \hat{W}_r & = & r^{-1} (\hat{W} - 1)(\hat{a}
\hat{W}
+ \hat{b}) + c \hat{W} \frac{1 + W}{1 - W}.
\end{eqnarray}
\end{mathletters}Here  $W$ and $\hat{W}$ satisfy eq.(\ref{P5}) with 
the
coefficients
($\alpha,\beta,\gamma,\delta$) and ($\hat{\alpha},\hat{\beta},
\hat{\gamma},\delta$), respectively, where $\hat{a}^2 = 2
\hat{\alpha}$,
$\hat{b}^2 = - 2 \hat{\beta}$, $\hat{\gamma} = c (b-a)$, and
$ 2\hat{a}  =   a+b-1 -\gamma/c$,
$2\hat{b}  =    a+b-1+ \gamma/c$.
With $\alpha$, $\beta$ and $\gamma$ as in eq.(\ref{coefficients}),
eqs.(\ref{bel2})
amount to the vorticity-raising transformations
(\ref{sys}).

We conclude the discussion of the complex sine-Gordon-1 equation by
mentioning that it would be natural to expect its vortex solutions
(confined to a finite region on the plane) to arise as degenerate
cases
of its $N$-soliton solutions \cite{BG2} (which have the form of $N$
intersecting infinite folds). This kind of correspondence between
two-dimensionally localized ``lumps" and one-dimensional
multisolitons
exists, for example, in the Kadomtsev-Petviashvili equation
\cite{Its}.
Surprisingly, the only two-dimensionally localized bounded solution
resulting from the ``degeneration" of the generic two-soliton
solution of eq.(\ref{csG-I}) is discontinuous at the origin:
$\psi =  ( X^2 - \sinh^2 Y)(X^2 + \sinh^2 Y)^{-1}$.
Here $X+iY = e^{i \alpha} (x+iy)$, and  $\alpha$ is an arbitrary
constant
angle.

{\it Vortices of the complex sine-Gordon-2.}  The
complex sine-Gordon-2  results from the variation of
eq.(\ref{E_cSG2}):
\begin{equation}
\nabla^2 \psi + \frac{(\nabla \psi)^2 \, {\overline \psi}}{2 - 
|\psi|^2}
+ \frac{1}{2} \psi(1-|\psi|^2) (2 - |\psi|^2) =0.
\label{csG-II}
\end{equation}
The multivortex Ansatz $\psi = \Phi_n(r) e^{i n \theta}=
{Q_n^{1/2}(r)} e^{i n \theta}$ takes it to
\begin{eqnarray}
\nonumber \frac{d^2 Q_n}{d r^2} + \frac{1}{r} \frac{d Q_n}{d r} & + &
\frac{1-Q_n}{Q_n(Q_n-2)} \left( \frac{d Q_n}{dr} \right)^2 +
Q_n \left( 1 - Q_n \right) \left( 2 - Q_n \right) \\ \label{ODE2}
& + & \frac{(a_n^2-b_n^2) Q_n}{r^2 (2 - Q_n)} + \frac{4 a_n^2
(1 - Q_n)}{ r^2 Q_n (2-Q_n)} = 0,
\end{eqnarray}
where $a_n = 0$ and $b_n = -2n$. Next, the substitution
$$
Q_n =  2 (1 - W)^{-1}
$$
transforms eq.(\ref{ODE2}) to the Painlev\'e  equation (\ref{P5})
with coefficients
$$
\alpha = 0,\;\;\;\;\beta = - 2 n^2,\;\;\;\;\gamma = 0,\;\;\;\;\delta
= 2.
$$ This time,
in order to construct the multivortex solutions we apply the
Schlesinger transformation (\ref{bel2}) twice.
This  leads to a recurrent
relation
\begin{equation}
\label{Sch}
Q^{(k-1)} = (2 - Q^{(k)}) \left\{ 1 - \frac{2 (a_k+b_k - 1) Q^{(k)}
\left[r Q^{(k)}_r - 2 a_k + (a_k+b_k) Q^{(k)} \right]}{
\left[r Q^{(k)}_r - 2 a_k + (a_k+b_k) Q^{(k)} \right]^2 +
r^2 Q^{(k)2} (2 - Q^{(k)})^2} \right\},
\end{equation}
where $Q^{(k)}$ and $Q^{(k-1)}$ satisfy eq.(\ref{ODE2}) with the
parameters
($a_k,\,b_k$) and ($a_{k-1},b_{k-1}$), respectively. Here $a_{k-1} =
a_k-1$ and $b_{k-1}=b_k-1$. Starting with a trivial solution 
$Q^{(0)}=1$
arising for $a_0 = - b_0 = n$, and using
eq.(\ref{Sch}) $n$ times, we end up with a solution
$Q_n=Q^{(-n)}$
which satisfies eq.(\ref{ODE2}) with $a_n = 0$ and $b_n = -2n$ and
the
boundary condition $Q_n \rightarrow 1$ as $r \rightarrow \infty$. 
These
 solutions are given by rational functions; in particular,
the first three multivortices (see Fig.1) read
\begin{eqnarray*}
Q_1 & = & \frac{r^2}{r^2 +4}; \\
Q_2 & = & \frac{r^4 (r^2 +24)^2}{r^8 +64r^6 +1152r^4 +9216r^2
+36864}; \\
Q_3 & = & r^6 (r^6 +144r^4 +5760r^2 + 92160)^2 {D_3}^{-1},\\
 D_3 & = & r^{18} +324 r^{16} + 41472 r^{14} + 2820096 r^{12}
+114130944 r^{10}   + 2919628800 r^8 \\ & +& 50960793600 r^6 +
611529523200 r^4 + 4892236185600 r^2 + 19568944742400.
\end{eqnarray*}
The energy of the complex sine-Gordon-2 vortices is logarithmically 
divergent.

{\it Concluding remarks.\/}
The Ginsburg-Landau expansion (\ref{GL}) is
regarded as a central
 postulate in the phenomenological theory of phase
transitions; however,
for some systems
eqs.(\ref{E_cSG1})-(\ref{E_cSG2})
may happen to provide a more adequate description.
In fact, the difference is not as big as one might think.
Assuming, for instance,  $|\psi|^2 \leq 1$,
eq.(\ref{E_cSG2}) can be rewritten as
\begin{equation}
\label{a1}
E_{SG-2} \approx \int \left[ |\nabla \psi|^2 + \frac{1}{2} (1 - 
|\psi|^2)^2
+
\frac{ |\nabla \psi|^2 |\psi|^2}{2} + ...\right]{d^2x};
\end{equation}
this is  different from (\ref{GL}) only in the third term which
is small both when $\psi \sim 0$ and when
$|\psi| \sim 1, \nabla \psi \sim 0$.
More importantly, the complex sine-Gordon models provide a unique
opportunity for studying a number of analytic properties
which are common to a wide class of vortex-bearing systems.
These include the correct Ansatz for two spatially separated
vortices, the vortex-phonon scattering matrix and so on;
our present construction of coaxial multivortices
 is hopefully but a first step in this direction. Finally, one may see
 the complex
sine-Gordon vortices as a starting point in the
{\it perturbative} construction of the corresponding solutions
of the Ginsburg-Landau and ferromagnet models.

 We are grateful to M. Bogdan, B. Dubrovin,
 B. Ivanov, A. Kapaev, P. Winternitz
and the referee
for useful remarks, and to N. Alexeeva and A. Harin for their
numerical assistance at various stages of this work.
One of the authors (I.B.) thanks S. Randjbar-Daemi for the
hospitality at ICTP.
This research was supported by the FRD of South Africa and
URC of UCT.

\newpage
\begin{center}
{\Large Figure Caption}
\end{center}
{\bf Fig.1} \, The vortex solutions with $n=1,2$ and $3$.

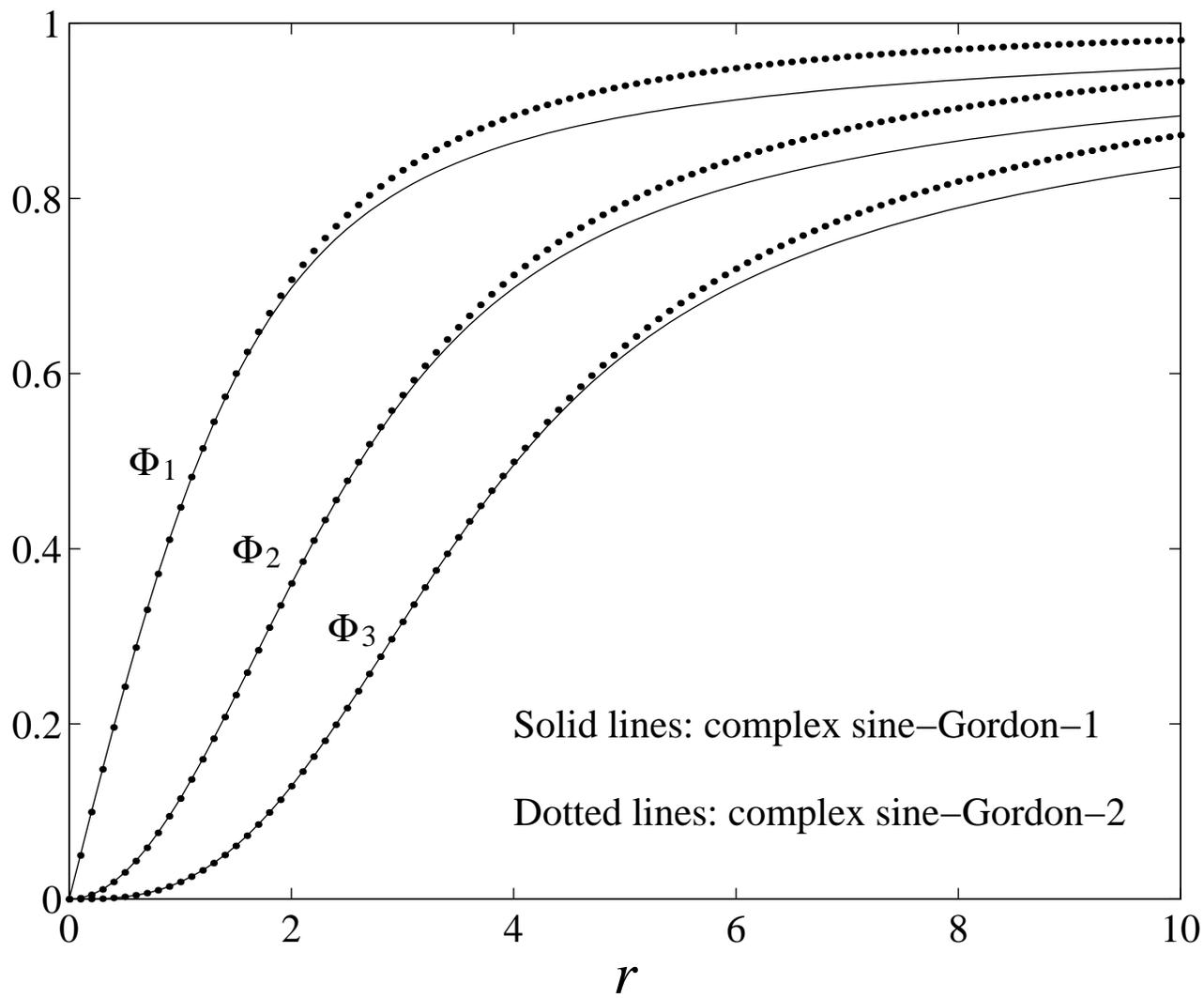
\end{document}